\documentclass[a4paper,11pt]{article}
\pdfoutput=1 

\usepackage{jcappub} 
\usepackage[T1]{fontenc} 
\usepackage{epsfig}
\usepackage{graphicx}
\usepackage{amsfonts}
\usepackage[usenames,dvipsnames]{xcolor}
\usepackage{url}
\usepackage{subfigure}
\usepackage{times,graphicx,amsmath,amsfonts,amssymb,epstopdf,hyperref}
\usepackage[normalem]{ulem}
\usepackage{multirow}

\def\ie{{\frenchspacing\it i.e.}}
\def\eg{{\frenchspacing\it e.g.}}

\def\be{\begin{equation}}
\def\ee{\end{equation}}
\def\ba{\begin{eqnarray}}
\def\ea{\end{eqnarray}}

\title{A model independent constraint on the temporal evolution of the speed of light}

\author[a,b]{Dandan Wang,}
\author[c]{Hanyu Zhang,}
\author[a,b]{Jinglan Zheng,}
\author[a]{Yuting Wang,}
\author[a,d,e]{Gong-Bo Zhao}

\affiliation[a]{National Astronomical Observatories, Chinese Academy of Science, Beijing, 100101, P. R. China}
\affiliation[b]{University of Chinese Academy of Science, Beijing, 100049, P. R. China}
\affiliation[c]{Department of Physics, Kansas State University, 116 Cardwell Hall, Manhattan, KS 66506, USA}
\affiliation[d]{School of Astronomy and Space Science, University of Chinese Academy of Sciences, Beijing, 100049, P. R. China}
\affiliation[e]{Institute of Cosmology \& Gravitation, Dennis Sciama Building, University of Portsmouth, Portsmouth, PO1 3FX, UK}

\emailAdd{ddwang@nao.cas.cn}
\emailAdd{hanyuz@phys.ksu.edu}
\emailAdd{jlzheng@nao.cas.cn}
\emailAdd{ytwang@nao.cas.cn}
\emailAdd{gbzhao@nao.cas.cn}

\abstract{We present a new, model-independent method to reconstruct the temporal evolution of the speed of light $c(z)$ using astronomical observations. After validating our pipeline using mock datasets, we apply our method to the latest BAO and supernovae observations, and reconstruct $c(z)$ in the redshift range of $z\in[0,1.5]$. We find no evidence of a varying speed of light, although we see some interesting features of $\Delta c(z)$, the fractional difference between $c(z)$ and $c_0$ (the speed of light in SI), \eg, $\Delta c(z)<0$ and $\Delta c(z)>0$ at $0.2\lesssim z\lesssim0.5$ and $0.8\lesssim z\lesssim1.3$, respectively, although the significance of these features is currently far below statistical importance.}

\begin{document}
\maketitle
\flushbottom

\section{Introduction}
\label{sec:intro}

In this era of precision cosmology, we are fortunately equipped with high quality observational data of various kinds, which makes it possible to test fundamental laws of the Universe. Physical laws of the Universe are built with a few ``constants'', including the gravitational constant $G$, the electron charge $e$, the speed of light $c$, and so on. Actually, assuming the constancy of these ``constants'' at all cosmic times and scales is a significant extrapolation of our knowledge on a rather limited temporal and spacial scales, which is subject to observational scrutiny.

Theorists including Dirac started thinking about building models for a varying $G$ or $e$, well before any observational tests became feasible \cite{Dirac,e}. However, the constancy of $c$ was much more sacred \cite{Barrow99}, as it is the pillar of special relativity, built by Einstein in 1905. Making $c$ vary is much more destructive to the structure of formalisms in modern physics, than varying other constants may do. Nevertheless, varying speed of light (VSL) may solve the cosmological constant problem and build a new framework of cosmic structure formation, as an alternative to inflation. This is sufficiently attractive for theorists to develop viable VSL theories (see \cite{VSLreview} for a review of recent developments in VSL). On the other hand, techniques for testing VSL observationally have been developed in parallel. Recently, a new method to constrain VSL at a single redshift using baryonic acoustic oscillations (BAO) was developed \cite{VSLBAO}, and has been applied to cosmological observations \cite{Cao,Cai}.

The method developed in \cite{VSLBAO} is briefly summerised as follows. Given the relation between the angular diameter distance $D_{A}(z)$ and the Hubble function $H(z)$, the speed of light $c$ at a specific redshift $z=z_{\rm M}$ is just a product of 
$D_A(z_{\rm{M}})$ and $H(z_{\rm{M}})$, where $z_{\rm M}$ is the peak location of $D_{A}(z)$, \ie, $\partial D_{A}/\partial z | _{z=z_{\rm M}}=0$. Note that the exact value of $z_{\rm M}$ is model-dependent, and $z_{\rm M}\sim1.7$ for a flat $\Lambda$CDM model that is consistent with the Planck {\it 2018} observations \cite{PLC18}.

This method has advantages of being model-independent and free from degeneracy with other cosmological parameters, but it has drawbacks. First, it is difficult to determine $z_{\rm{M}}$ accurately, as $D_{A}(z)$ is rather flat around its peak. Second, for a wide range of cosmologies, $z_{\rm M}$ is as large as $\sim1.7$, making it unaccessible for most current BAO observations. Even for future deep BAO surveys including Dark Energy Spectroscopic Instrument (DESI) \cite{DESI}, it is challenging to get decent $D_{A}$ or $H$ measurements at such high redshifts. Last but not least, this method only constrains $c$ at one single redshift, which is far from sufficient for tests against various VSL models.

In this work, we generalise \cite{VSLBAO} to propose a new model-independent method for testing VSL at multiple redshifts. After presenting the methodology in Sec. \ref{sec:method}, we perform validation tests using mock datasets, before applying to observational data and presenting our main result in Sec. \ref{sec:realdata}. We conclude and discuss our results in Sec. \ref{sec:conclusion}.

\section{Methodology}
\label{sec:method}

We start from the relation between the angular diameter distance $D_{A}(z)$, the Hubble function $H(z)$, and the general speed of light function $c(z)$ in a flat FRW Universe \footnote{Note that in non-flat Universes, the Friedmann equations get modified by the time derivative of $c(z)$ so Eq ({\ref{eq:DAH}}) does not hold \cite{Barrow98}. In this work, we consider a flat Universe for simplicity.}, \ie,
\ba \label{eq:DAH} D_{A}(z)=\frac{1}{1+z} \int_{0}^z \frac{c(z')}{H(z')} {\rm d} z',  \ea Differentiating both sides with respect to redshift $z$ yields,
\be \label{eq:c} c(z)=\chi'(z) H(z),  \ee where $\chi$ is the comoving distance so that $\chi(z)=(1+z)D_{A}(z)$. Our aim is to reconstruct the entire evolution history of $c(z)$, thus we parametrise the redshift-dependence of $\chi(z)$ and $H(z)$ as follows,

\ba \label{eq:chi} \frac{\chi(z)}{\chi_{\rm{fid}}(z)}&=&\alpha_{0}\left(1+\alpha_{1}x+\frac{1}{2}\alpha_{2}x^2+\frac{1}{6}\alpha_{3}x^3\right), \\
\label{eq:H} \frac{H_{\rm{fid}}(z)}{H(z)}&=&\beta_{0}+\beta_{1}x+\frac{1}{2}\beta_{2}x^2+\frac{1}{6}\beta_{3}x^3. \ea The variable $x\equiv\chi_{\rm{fid}}(z)/\chi_{\rm{fid}}(z_{\rm{p}})-1$ where the subscript $_{\rm{fid}}$ denotes the fiducial cosmology used, and $z_{\rm{p}}$ is the pivot redshift at which we apply the Taylor expansion. In this work, the fiducial cosmology is chosen to be a flat $\Lambda$CDM model with $\Omega_{\rm{M,fid}}=0.31$.

To test against the constancy of the speed of light, we define a deviation function of $c$ as, \ba\label{eq:dc} \Delta c(z) \equiv \frac{c(z)}{c_0}-1, \ea where $c_0$ is the speed of light in the International System of Units (SI), \ie, $c_0 = 299,792,458 \ {\rm m \ s}^{-1}$.

Combining Eqs (\ref{eq:c}), (\ref{eq:chi}), (\ref{eq:H}) and (\ref{eq:dc}), we have,
\be \label{eq:dcz} \Delta c(z)=\frac{\alpha_{0}\left[1+\alpha_{1}+\left(2\alpha_{1}+\alpha_{2}\right)x+\left(\frac{3}{2}\alpha_{2}+\frac{1}{2}\alpha_{3}\right)x^2\right]}{\beta_{0}+\beta_{1}x+\frac{1}{2}\beta_{2}x^2.} -1.\ee Specially, at $z_{\rm{p}}$ where $x$ vanishes,
\be \label{cbarzp} \Delta c(z_{\rm{p}})=\frac{\alpha_{0}(1+\alpha_{1})}{\beta_{0}} -1.\ee

Although $\Delta c$ does not explicitly depends on other cosmological parameters, its measurement does implicitly rely on how well we are able to model $D_A$ and $H$ using Eqs (\ref{eq:chi}), (\ref{eq:H}) for general cosmologies. The expansion Eq (\ref{eq:chi}) was actually proposed by \cite{Zhu14} for implementing the optimal redshift weighting method for BAO analyses, and it was shown that expansions up to the quadratic order in $x$ can precisely recover a wide range of cosmologies at the sub-percent level in the redshift range of $z\in[0.5,1.5]$. Note that in \cite{Zhu14}, the speed of light was assumed to be a constant, thus $H(z)$ was derived from $\chi(z)$. In our case, as $c(z)$ is promoted to a general function of $z$, we have to double the number of the expansion coefficients for $H(z)$. Also, we are more ambitious on the valid redshift range for these expansions, since we plan to use all the available observations probing the background expansion of the Universe from $z=0$ to $z=2$. All these motivated us to use expansions to higher order to achieve the precision we need.

To validate our parametrisations for $\chi(z)$ and $H(z)$, we attempt to fit two cosmologies which sufficiently deviate from the fiducial cosmology we use for the expansion. Specifically, 
\ba
\label{eq:cosmo1}&& {\rm Cosmo. \ Model \ I: \ a \ flat \ \Lambda CDM \ model \ with} \ \Omega_{\rm M}=0.2;  \\
\label{eq:cosmo2}&& {\rm Cosmo. \ Model \ II: \ a \ flat } \ w{\rm CDM \ model \ with} \ \Omega_{\rm M}=0.31; \ w=-0.8.
\ea
Both models are excluded by Planck {\it 2018} observations, making them representative for ``extreme'' models that may be later sampled in the parameter space. For each model, we first choose a pivot redshift $z_{\rm p}$, and then tune the coefficients $\alpha$'s and $\beta$'s to minimise the difference between our model prediction for $D_A$ and $H$ computed using Eqs (\ref{eq:chi}), (\ref{eq:H}), and the exact quantities calculated using models I and II. Then we compute the residual $R$ for $D_A(z)$ and $H(z)$ respectively, \ie,
\ba\label{eq:R} R_o(z,z_{\rm p}) \equiv 100\times \left[\frac{O^{\rm model}(z,z_{\rm p})}{O^{\rm exact}(z,z_{\rm p})}-1\right]\ \%, \ea where $O$ stands for $D_A$ or $H$.

\begin{figure}
\centering
{\includegraphics[scale=0.6]{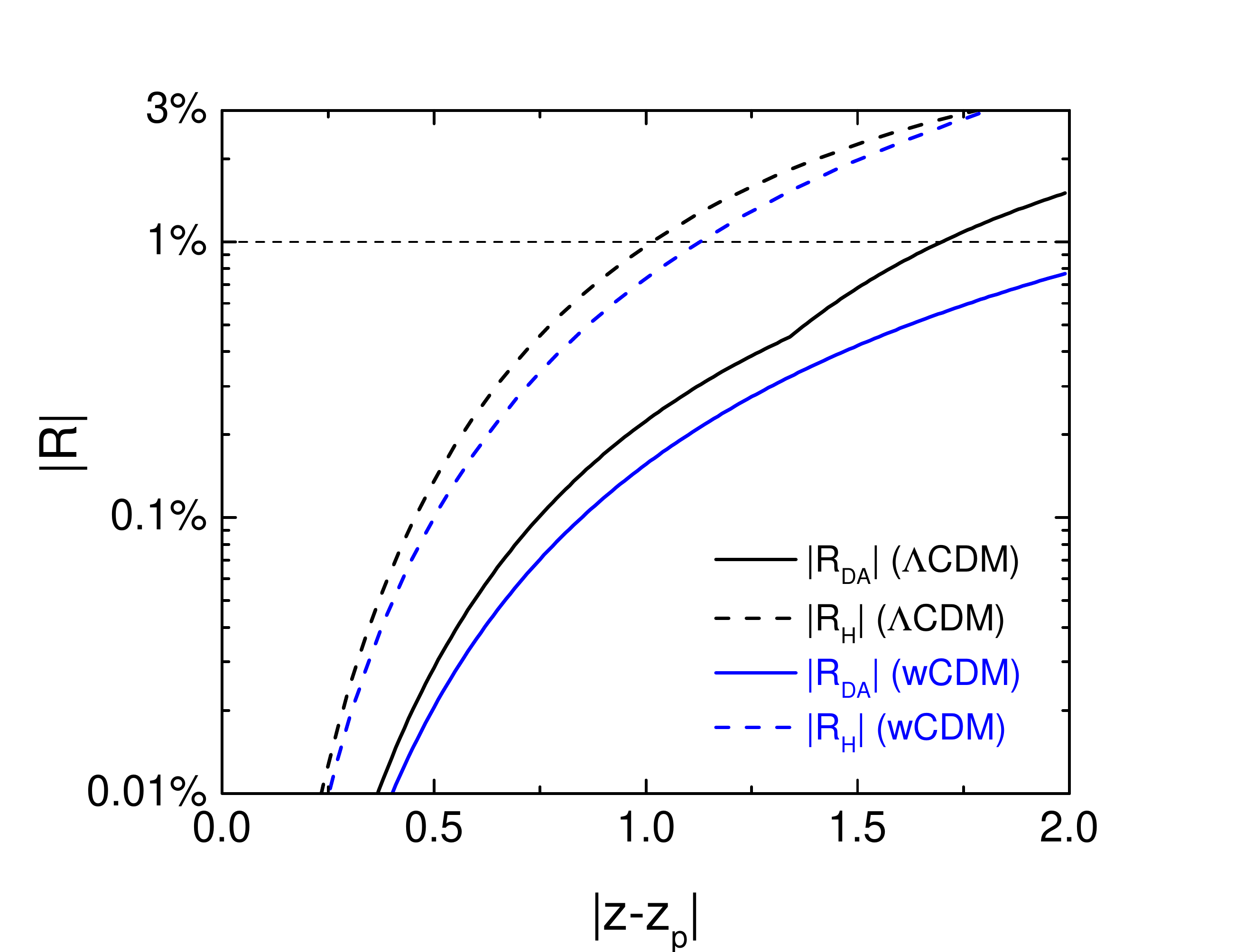}}
\caption{The residual $R$ defined in Eq (\ref{eq:R}) as a function of $|z-z_{\rm p}|$. As shown, the residuals are below the $1\%$ level for $|z-z_{\rm p}|<1$, although $H(z)$ is less accurately modeled than $D_A(z)$. For larger $|z-z_{\rm p}|$, say, $1<|z-z_{\rm p}|<2$, $R$ for $D_A(z)$ can still be controlled at the per cent level, while the residual for $H$ approaches the 4\% level.}
\label{fig:precision}
\end{figure}

The computed residuals are shown in Fig. \ref{fig:precision} for various $|z-z_{\rm p}|$ values, and we find that our model with third-order expansions are sufficiently accurate for our analysis: the residuals in most cases are below $1\%$ level, although $H(z)$ is less accurately modeled than $D_A(z)$. Larger residuals appear when $|z-z_{\rm p}|$ gets larger, which is naturally expected as the accuracy of the Taylor expansion decays with the distance to the expansion point. Quantitatively, $|R|\lesssim 1\%$ for $|z-z_{\rm p}|\lesssim 1$ for both models, and $|R|$ can approach $4\%$ for $H$ in the worst case, \eg, $|z-z_{\rm p}|=2$, but as we will explain later, this does not affect our result, if we only take the reconstructed values at $z_{\rm p}$ (so that $|z-z_{\rm p}|=0$). 

Before applying our method to actual observations, there is another step missing: an actual mock test for various VSL models, which is presented in the next section.

\section{Mock tests}
\label{sec:mock}

This section is devoted to a robustness test of our pipeline, before applying to actual observations presented in the next section.

To begin with, we choose four phenomenological VSL models for $\Delta c(z)$ to cover various possible features, including a constant, a linear function, quadratic function and an oscillatory function, \ie,
\begin{itemize}
\item VSL model 1: $\Delta c(z)=0.05$;
\item VSL model 2: $\Delta c(z)=0.05z$;
\item VSL model 3: $\Delta c(z)=0.05z+0.01z^2$;
\item VSL model 4: $\Delta c(z)=0.05\sin(z\pi)$.
\end{itemize}

These VSL models are then combined with two cosmological models, respectively, as shown in Eqs (\ref{eq:cosmo1}) and (\ref{eq:cosmo2}) to make eight toy models for the mock test (see Table \ref{tab:mock} for the labelling of the toy models).

For each toy model, we can first generate mock $H(z)$ data points using cosmological models shown in Eqs (\ref{eq:cosmo1}) and (\ref{eq:cosmo2}), and then produce mock $D_A(z)$ datasets by combining $H(z)$ and $\Delta c(z)$ using Eq (\ref{eq:DAH}).

In practice, we produce two kinds of mock datasets, a combined BAO data sample assuming the sensitivity of the Euclid mission \citep{Euclid} and DESI survey \citep{DESI}, and a combined supernovae sample forecasted for LSST \citep{LSST} and Euclid \citep{EuclidSNIa}.

For the BAO sample, we follow \citep{Euclid} and \citep{DESI} to produce $D_{A}$ and $H$ pairs at $15$ and $18$ effective redshifts for Euclid and DESI, respectively, so that our combined BAO sample consists of $33$ pairs of $D_{A}$ and $H$ covering the redshift range of $z\in(0,2.1)$. For the supernovae sample, we produce luminosity distances assuming the sensitivity of the Deep-Drilling Fields and low-redshift sample to be observed by LSST, and of the DESIRE supernovae survey of Euclid. The Deep-Drilling Fields will observe $8800$ supernovae at $z\in[0.15,0.95]$, and additional $8000$ supernovae below redshift $0.35$. This sample is complemented by the high-$z$ sample to be collected by the DESIRE survey, which will provide $1740$ supernovae at $z\in[0.75,1.55]$.

\begin{table}
\begin{center}
\begin{tabular}{c| c| c}
\hline\hline
Toy model&VSL model&Cosmological model\\
\hline
1&VSL model 1&$\Lambda$CDM; Eq (\ref{eq:cosmo1})\\ \hline
2&VSL model 2&$\Lambda$CDM; Eq (\ref{eq:cosmo1})\\ \hline
3&VSL model 3&$\Lambda$CDM; Eq (\ref{eq:cosmo1})\\ \hline
4&VSL model 4&$\Lambda$CDM; Eq (\ref{eq:cosmo1})\\ \hline
5&VSL model 1&$w$CDM; Eq (\ref{eq:cosmo2})\\ \hline
6&VSL model 2&$w$CDM; Eq (\ref{eq:cosmo2})\\ \hline
7&VSL model 3&$w$CDM; Eq (\ref{eq:cosmo2})\\ \hline
8&VSL model 4&$w$CDM; Eq (\ref{eq:cosmo2})\\ \hline
\end{tabular}
\caption{Toy models used for the mock test.}
\label{tab:mock}
\end{center}
\end{table}

\begin{figure}
\centering
{\includegraphics[scale=0.2]{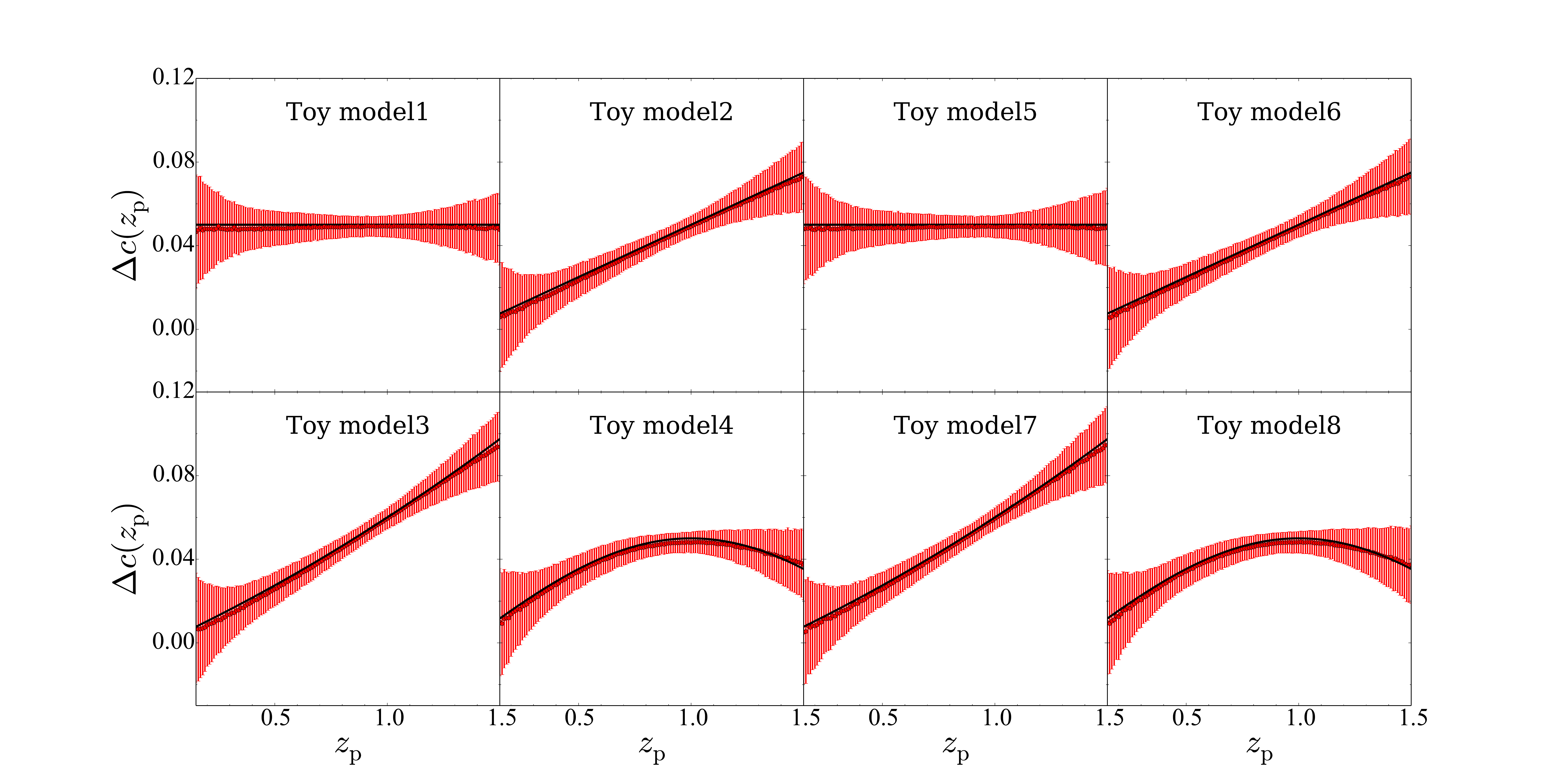}}
\caption{The result of the mock test for eight toy models shown in Table \ref{tab:mock}. The black solid lines and the red dots show the input and reconstructed values of $\Delta c(z_{\rm p})$, respectively, and the error bars illustrate the 68\% CL uncertainty.}
\label{fig:mock}
\end{figure}

For a given $z_{\rm p}$, these datasets can be fitted with theoretical models Eqs (\ref{eq:chi}) and (\ref{eq:H}) for parameters $\alpha$'s and $\beta$'s, using a modified version of {\tt CosmoMC} \cite{cosmomc}. We could then use the resultant $\alpha$'s and $\beta$'s to reconstruct $\Delta c(z)$ using Eq (\ref{eq:dcz}). However, this result may be subject to systematic errors in $\Delta c(z)$ at redshifts that are far away from $z_{\rm p}$, as we have seen in Fig. \ref{fig:precision}. It is true that the residual $|R|$ is as low as 3\% in the worst case as discussed in Sec. \ref{sec:method}, but it can be further improved.

The way out is to abandon the reconstructed $\Delta c(z)$ at all redshifts except for $z=z_{\rm p}$, where the Taylor expansion is error free, and repeat the fitting process for every $z_{\rm p}$ (with equal space of $\Delta z=0.01$ in $z$) running in the entire redshift range. This is computationally expensive, and the error of $\Delta c(z)$ at different redshifts are highly correlated, but it is much more accurate.

The result of this mock test is summerised in Fig. \ref{fig:mock}. As shown, the input VSL models are reconstructed perfectly for all toy models, with negligible bias compared to the statistical error budget.

\section{Implication on the latest observational datasets}
\label{sec:realdata}

Now it is time to apply our pipeline for measuring $\Delta c(z)$ to actual observational data. We shall first introduce datasets we use, and then present the result.

\subsection{The observational dataset}
\label{sec:data}
The datasets used in this analysis include,
\begin{itemize}
\item {\bf The BAO measurements.} As we are interested in reconstructing the time evolution of $\Delta c$, we use the tomographic BAO measurements from the Baryonic Oscillation Spectroscopic Survey (BOSS) Data Release (DR) 12 sample, which provides measurement of $D_A/r_{\rm d}$ and $H r_{\rm d}$ pairs at nine effective redshifts in the redshift range of $z\in[0.31,0.64]$ \cite{DR129bin}. To approach the high-$z$ end, we also use the tomographic BAO measurement from the extended BOSS (eBOSS) DR14 quasar sample, which offered four additional $D_A/r_{\rm d}$ and $H r_{\rm d}$ pairs at redshifts $z\in[0.98,1.94]$ \cite{DR144bin}. Note that as Eq (\ref{eq:c}) shows, the speed of light is a product between $H$ and ${\rm d}\chi/ {\rm d} z$, $r_{\rm d}$ cancels out.

\item {\bf The observational $H(z)$ data (OHD).} The OHD, as `cosmic chronometers', are measured using the ages of passively evolving galaxies, and we use a compilation of $30$ data points shown listed in Table \ref{ohd}.

\item {\bf The supernovae data.} We use the `joint light-curve analysis (JLA)' sample \cite{JLA}, which is a re-anaylsis of $740$ data points consisting of samples from the three year SDSS-II supernovae survey \cite{Sako:2014qmj} and the SNLS samples presented in \cite{conley2011}. 

\end{itemize}

\subsection{The final result}
\label{sec:result}

\begin{figure}
\centering
{\includegraphics[scale=0.4]{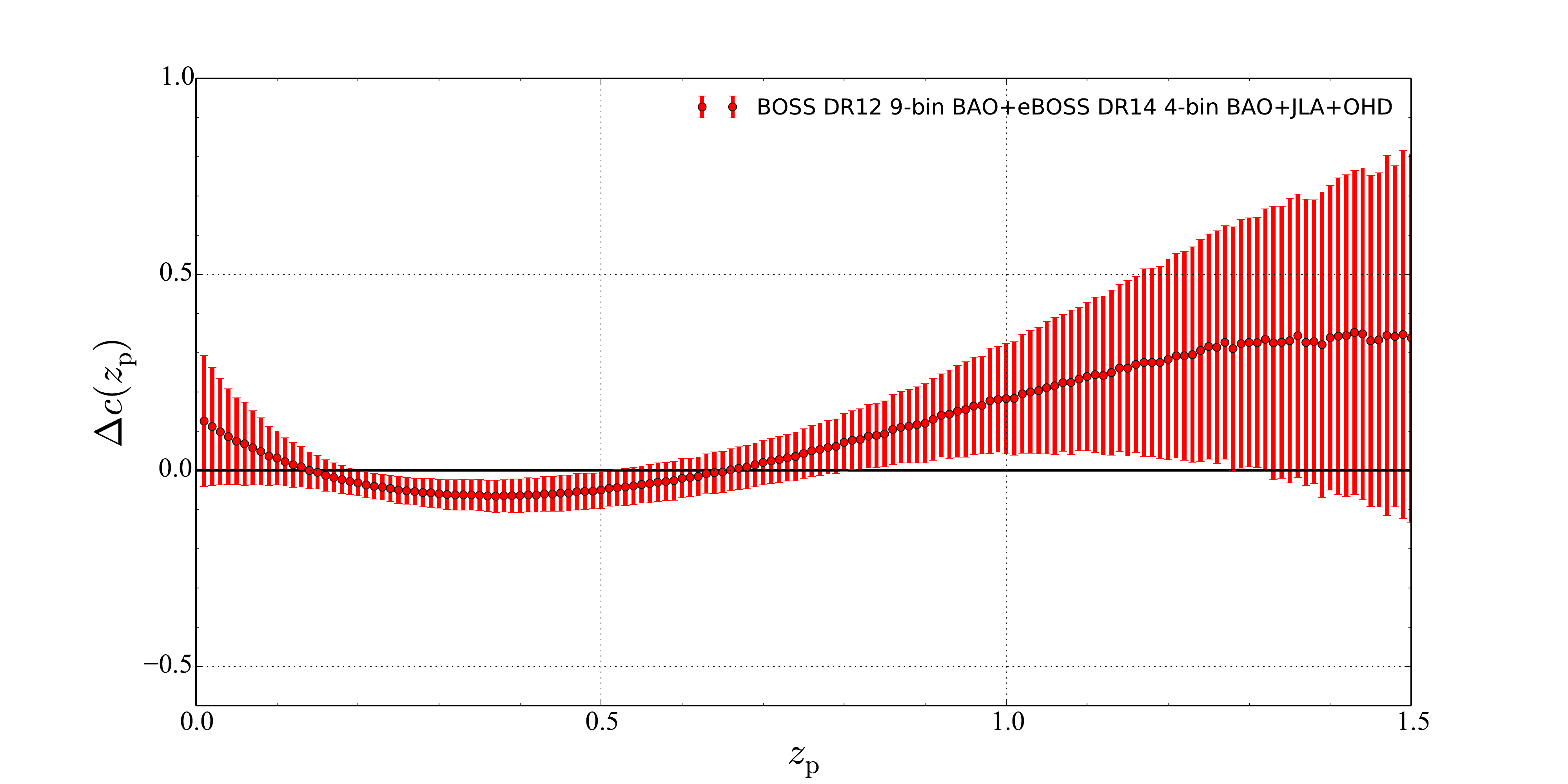}}
\caption{A reconstruction of $\Delta c(z_{\rm p})$ derived from a compilation of current observations described in Sec. \ref{sec:data}. The black horizontal line shows $\Delta c(z_{\rm p})=0$ to guide eyes.}
\label{fig:result}
\end{figure}

We apply our pipeline to the actual observational data, in the same way as we performed on the mock data, and present the result in Fig. \ref{fig:result}. Again, this is a compilation of $\Delta c(z_{\rm p})$ for numerous $z_{\rm p}$'s with equally spaced in $z$ with $\Delta z=0.01$.

As shown, the uncertainty of $\Delta c(z_{\rm p})$ gets minimised at $z\sim0.2$, where the Universe is best probed by current supernovae and BAO experiments. The constraints on the low-$z$ and high-$z$ ends are looser, because of the small volume at low-$z$ and the sparsity of galaxy distributions at high-$z$, which dilutes the BAO constraints. Furthermore, the current supernovae surveys can barely access the Universe at $z\gtrsim1$.

\begin{figure}
\centering
{\includegraphics[scale=0.4]{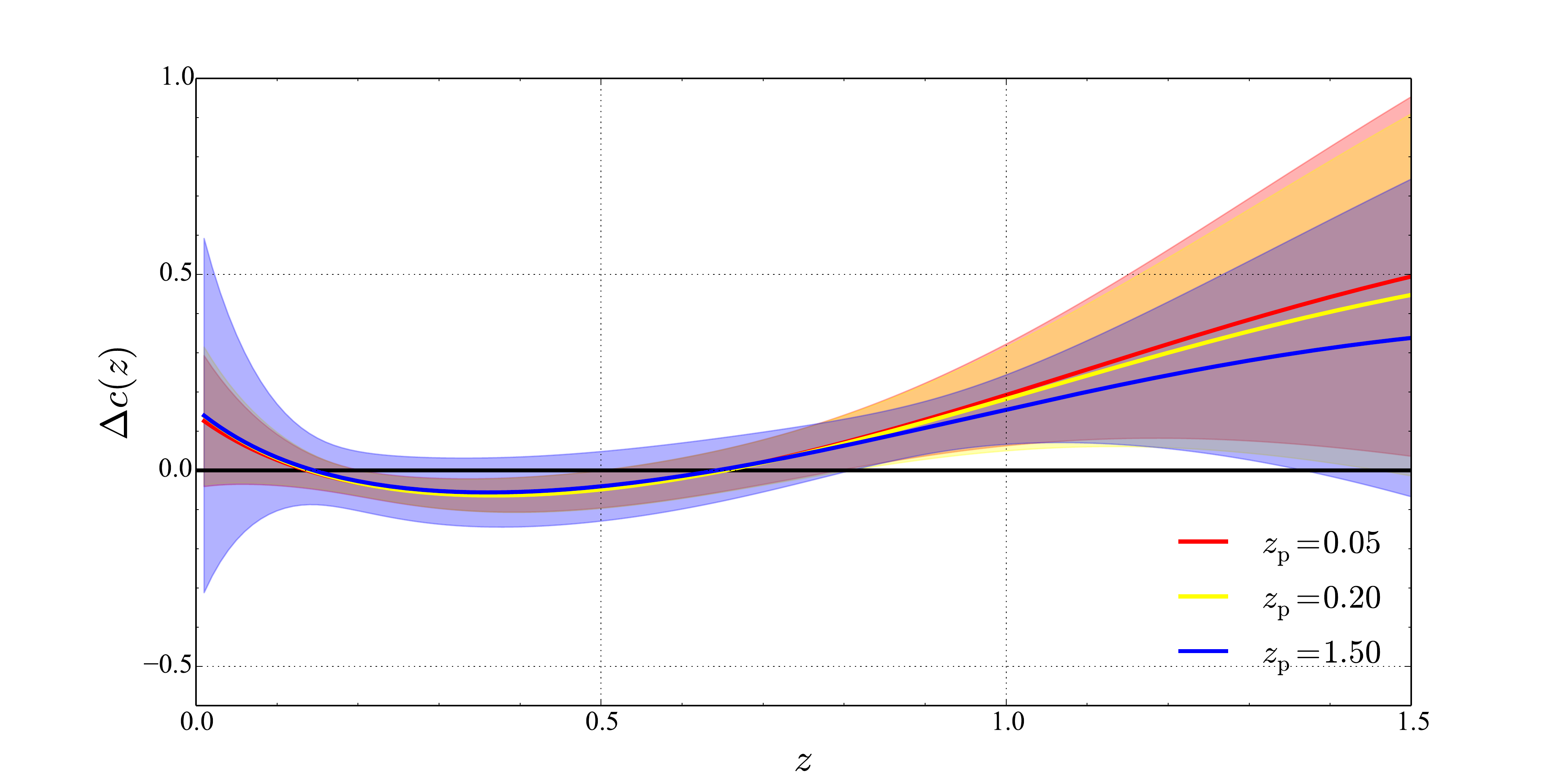}}
\caption{The reconstructed $\Delta c(z)$ using three values of $z_{\rm{p}}$, as illustrated in the legend. The shaded bands show the 68\% uncertainty.}
\label{fig:recover}
\end{figure}

The result is consistent with $\Delta c(z)=0$ given the level of uncertainty, but interesting features show up at $0.2\lesssim z\lesssim0.5$ and $0.8\lesssim z\lesssim1.3$ where $\Delta c(z)<0$ and $\Delta c(z)>0$, respectively, which awaits further investigation using future observations.

We are interested in quantifying the signal-to-noise ratio of $\Delta c(z)\ne0$, yet it is not straightforward because the error bars are highly correlated, but a covariance matrix is unavailable as we performed the parameter constraints at various $z_{\rm p}$ individually.

However, for a given $z_{\rm p}$, we are able to quantify the deviation of $\Delta c(z)$ from zero with the reconstructed $\Delta c(z)$ using Eq (\ref{eq:dcz}), as the covariance for the $\alpha$ and $\beta$ parameters is available. Before proceeding, we show the reconstructed $\Delta c(z)$ in Fig. \ref{fig:recover} using three values of $z_{\rm{p}}$ at $0.05, 0.2$ and $1.5$, which covers a reasonable choice of $z_{\rm p}$ ($0.2$) and two extreme values ($0.05$ and $1.5$). As shown, these results agree reasonably well with that shown in Fig. \ref{fig:result}, although the result using $z_{\rm p}=1.5$ over-estimates the uncertainty (but the central value remains accurate).

We then quantify the difference between $\Delta c(z)$ and zero using the $\alpha$'s and $\beta$'s derived using a specific $z_{\rm p}$. Note that $\Delta c(z)=0$ means that the numerator is identical to the denominator of the fraction in Eq (\ref{eq:dcz}), which translates into,

\ba \label{eq:D} \Delta_0&\equiv&\beta_{0}-\alpha_{0}(1+\alpha_{1})=0,  \nonumber \\
\Delta_1&\equiv&\beta_{1}-\alpha_{0}(2\alpha_{1}+\alpha_{2})=0, \nonumber\\
\Delta_2 &\equiv& \beta_{2}-\alpha_{0}(3\alpha_{2}+\alpha_{3})=0.  \ea

Given the measured $\alpha$'s and $\beta$'s and the corresponding covariance matrix, we can easily evaluate the following $\chi^2$, \ba \chi^2_{z_{\rm p}} = {\Delta}^T C_{\Delta} {\Delta} \ea where $\Delta\equiv\{\Delta_0,\Delta_1,\Delta_2\}^T$ and $C_{\Delta}$ is the derived covariance matrix for vector $\Delta$. We average over the $\chi^2$ for all the $z_{\rm p}$'s and compute the mean and variance, which gives,
\ba {\chi}^2  = 3.2\pm 1.1.\ea which is statistically expected for fitting a fixed value, $0$, to three data points ($\Delta_i, \ i=1,2,3$). The error of $\chi^2$ quantifies the fluctuation of the result using various $z_{\rm p}$, thus it can be viewed as systematics. So the conclusion is, the constant speed of light measured in SI is consistent with current cosmological observations.

\section{Conclusion and discussions}
\label{sec:conclusion}

As a competitor of inflation, viable VSL theories deserve serious investigation both theoretically and observationally. With the accumulation of high quality cosmological datasets, it is time to develop theoretical and numerical tools to put constraints on the VSL scenario.

In this work, we take a phenomenological approach to study the VSL models. We propose a new method to constrain the speed of light using observations. Compared to previous works, our method enables a reconstruction of the temporal evolution of the speed of light $c(z)$, without dependence on cosmological models. 

After validating our method and pipeline using mock datasets of BAO and supernovae, we apply our method to the latest astronomical observations including the anisotropic BAO measurements from BOSS (DR12) and eBOSS (DR14 quasar), and the JLA supernovae sample, and reconstruct $c(z)$ in the redshift range of $z\in[0,1.5]$. We find no evidence of a varying speed of light, although we see some interesting features of $\Delta c(z)$, the fractional difference between $c(z)$ and $c_0$ (the speed of light in SI), \eg, $\Delta c(z)<0$ and $\Delta c(z)>0$ at $0.2\lesssim z\lesssim0.5$ and $0.8\lesssim z\lesssim1.3$, respectively. Although the significance of these features is far below statistical importance, it is worth further investigations using future observations.

\acknowledgments

This work is supported by the National Key Basic Research and Development Program of China (No. 2018YFA0404503), National Basic Research Program of China (973 Program) (2015CB857004), by NSFC Grants 11720101004 and 11673025 and by a CAS Interdisciplinary Innovation Team Fellowship. This research used resources of the SCIAMA cluster supported by University of Portsmouth.

\appendix

\section{The $H(z)$ data}

We summerise the OHD data used in this work in the following table.

\begin{table}
\begin{center}
\begin{tabular}{c c c c}
\hline\hline
  $z$& $H(z)$ & $\sigma_{H(z)}$& Reference\\ \hline
\hline
$0.07 $ & $69.0$ & $19.6$& \cite{ohd-p4} \\
$0.09$ & $69$ & $12$ &\cite{ohd-p2}\\ 
$0.12$&$68.6$&$26.2$&\cite{ohd-p4} \\
$0.017$&$83$&$8$&\cite{ohd-p2} \\
$0.179$&$75$&$4$&\cite{ohd-p1} \\
$0.199$&$75$&$5$&\cite{ohd-p1} \\
$0.20$&$72.9$&$29.6$&\cite{ohd-p4} \\
$0.27$&$77$&$14$&\cite{ohd-p2} \\
$0.28$&$88.8$&$36.6$&\cite{ohd-p4} \\
$0.352$&$83$&$14$&\cite{ohd-p1} \\
$0.3802$&$83$&$13.5$&\cite{ohd-p6} \\
$0.4$&$95$&$17$&\cite{ohd-p2} \\
$0.4004$&$77$&$10.2$&\cite{ohd-p6} \\
$0.4247$&$87.1$&$11.2$&\cite{ohd-p6} \\
$0.44497$&$92.8$&$12.0$&\cite{ohd-p6} \\
$0.4783$&$80.9$&$9$&\cite{ohd-p6} \\
$0.48$&$97$&$62$&\cite{ohd-p3} \\
$0.583$&$104$&$13$&\cite{ohd-p1} \\
$0.68$&$92$&$8$&\cite{ohd-p1} \\
$0.781$&$105$&$12$&\cite{ohd-p1} \\
$0.875$&$125$&$17$&\cite{ohd-p1} \\
$0.88$&$90$&$40$&\cite{ohd-p3} \\
$0.9$&$117$&$23$&\cite{ohd-p2} \\
$1.037$&$154$&$20$&\cite{ohd-p1} \\
$1.3$&$168$&$17$&\cite{ohd-p2} \\
$1.363$&$160$&$33.6$&\cite{ohd-p5} \\
$1.43$&$177$&$18$&\cite{ohd-p2} \\
$1.53$&$140$&$14$&\cite{ohd-p2} \\
$1.75$&$202$&$40$&\cite{ohd-p2} \\
$1.965$&$186.5$&$50.4$&\cite{ohd-p5} \\
\hline
\end{tabular}
\caption{$H(z)$ measurements (in unit of [${\rm km} \ {\rm s}^{-1} \ {\rm Mpc}^{-1}$]) used in this work.}
\label{ohd}
\end{center}
\end{table}

\end{document}